\documentclass[a4paper,11pt]{article}
\pdfoutput=1 

\usepackage{cancel}
\usepackage{jheppub} 
\usepackage[T1]{fontenc} 
\usepackage{overpic}

\usepackage{color}
\usepackage{amsthm}
\usepackage{graphicx}
\usepackage{slashed}
\usepackage{amssymb}

\def\slashchar#1{\setbox0=\hbox{$#1$}           
   \dimen0=\wd0                                 
   \setbox1=\hbox{/} \dimen1=\wd1               
   \ifdim\dimen0>\dimen1                        
      \rlap{\hbox to \dimen0{\hfil/\hfil}}#1 
   \else                                        
      \rlap{\hbox to \dimen1{\hfil$#1$\hfil}}/                                    \fi}

\title{\boldmath Spatial Mass}

\author{Benjamin Nachman}
\author{and Ariel Schwartzman}

\affiliation{SLAC, Stanford University, 2575 Sand Hill Rd, Menlo Park,
  CA 94025, U.S.A.}

\emailAdd{bnachman@cern.ch, sch@slac.stanford.edu}

\abstract{
In analogy to the transverse mass constructed from two objects, we define the {\it spatial mass} constructed from the input objects 3-vector momenta.  This observable is insensitive to jet mass scale and resolution uncertainties when constructed from small-radius hadronic jets.  Thus it improves the effective resolution on multijet masses for searches and measurements in hadronic final states.  To illustrate the efficacy of the spatial mass, we consider a top quark mass measurement at the Large Hadron Collider (LHC) in the 3-jet final state.  The reduction in uncertainty comes with a negligible cost in sensitivity.  
}

\begin{document} 
\maketitle
\flushbottom

\section{Introduction}

Compared to their spatial momentum, the mass of most final state particles at the Large Hadron Collider (LHC) are negligible.  However, collections of collimated particles known as jets can have non-negligible mass due to the opening angles between particles.  There have been many successful, direct uses of large radius, $R\sim 1.0$, jet mass to differentiate heavy particle decays from Standard Model (SM) multijet processes~(see e.g.~\cite{fatjet1,fatjet2,fatjet3,fatjet4}).  In these searches, the large radius jets capture most of the boosted decay products of the massive new physics particles.  As such, the jet mass is a powerful discriminant as it is a proxy for the mass of the sought after new physics particles.

There are also many indirect uses of jet masses for small, $R\sim0.4$, radius jets.  In particular, the use of di-, tri-, and multi-jet invariant masses is ubiquitous at the LHC.  Countless searches for physics beyond the SM and precision measurements of SM parameters make use of the invariant mass of several small radius jets to isolate signal processes such as $t\rightarrow bW(\rightarrow qq'), W\rightarrow qq',H\rightarrow b\bar{b}$, and $Z'\rightarrow q\bar{q}$ (c.f.~\cite{c1,c2,c3,c4,c5,a1,a2,a3,a4,a5} for a small set of examples).   The mass of a system of particles is often dominated by the momenta and opening angles of the constituent four-vectors, but the small radius jet masses also play a role.  Just as jet $p_T$ needs to be calibrated in experimental searches (beyond the input object calibrations), the jet mass also needs to be calibrated.  This calibration has uncertainties\footnote{One potentially large contribution is due to the local fluctuations of additional $pp$ collisions per bunch crossings known as pileup.} which can give a non-negligible contribution to the overall systematic uncertainty in a particular analysis.  The purpose of this paper is to illustrate the impact of small radius jet mass uncertainties on a physics measurement. We contrast this with a modified jet mass definition, the ``spatial mass.''  We will show that there is a tradeoff between the additional uncertainty from the full invariant mass and the loss of information in spatial mass.  Section 2 defines the spatial mass in an analogy to the transverse mass.  Section 3 introduces the top quark mass measurement, which will be used as an illustration throughout the paper.  Section 4 demonstrates the impact of small radius jet mass uncertainties on the top quark mass measurement.  Section 5 shows the results with the spatial mass variable.  Finally, Section 6 presents conclusions.  

\section{Spatial Mass}

Hadron colliders provide an exciting and challenging setting for probing the elementary nature of matter.   One of the key differences compared to a lepton collider is that the center of mass energy of a particular collision is not known.  Since the initial state has zero momentum transverse to the beam, many techniques have been developed to capitalize on the conservation of momentum in this plane.   For example, one powerful variable for isolating leptonic $W$ boson decays is the transverse mass.  Given two four-vectors $p^\mu$ and $q^\mu$, the transverse mass is defined as $m_T^2\equiv (f(p^\mu)+f(q^\mu))^2$, where $f$ maps: 

$$p^\mu=(p_x,p_y,p_z,E)\rightarrow f(p^\mu)=\left(p_x,p_y,0,\sqrt{p_x^2+p_y^2+m^2}\right),$$

\noindent i.e. $p_z\rightarrow 0$.  The insensitivity of $m_T$ to $p_z$ and also the associated calibration uncertainties is important because there is no conservation law in the longitudinal dimension.  Since the initial longitudinal momentum is not known, there is no way to calibrate $p_z$ {\it in situ}.  

The transverse mass is often used to isolate $W\rightarrow l+\nu$ events where the neutrino is not detected, but conservation of momentum in the transverse plane allows for its $\vec{p}_T$ to be inferred.  Another way of thinking about $m_T$ is as a variable which has reduced uncertainty  because it is insensitive to the uncertainty in the longitudinal direction of the momentum.  This is irrelevant for neutrinos, but consider $W\rightarrow j_1j_2$ for jets $j_1$ and $j_2$.  In this case, $m_T(j_1j_2)$ will have less uncertainty than $m_{j_1j_2}$, but will also contain less information.  The longitudinal momentum often contains useful information about the initial state so it is most likely ill advised to use $m_T$ in this case.  Furthermore, the lack of a conservation law in the $z$ direction is not a large concern because the transverse momentum magnitude can be measured well (and calibrated {\it in situ}) and the longitudinal angular resolution is usually subdominant, so the uncertainty on the scale and resolution of $p_z=p_T\sinh\eta$ is well understood.  

However, in addition to the lack of longitudinal conservation of momentum, there is also a missing constraint from the conservation of energy.  This fourth component of the input object four-vector cannot be inferred from the angular coordinates and $p_T$.  Instead, it depends on the object mass.  For hadronic jets, it is possible to measure all the components of the momentum, but the uncertainty in the energy is still an issue.  To circumvent the dependance on the jet mass, one may define, analogously to $m_T$, the spatial mass $m_s^2\equiv (g(p^\mu)+g(q^\mu))^2$, where $g$ is the function which maps:

$$p^\mu=(p_x,p_y,p_z,E)\rightarrow g(p^\mu)=\left(p_x,p_y,p_z,\sqrt{p_x^2+p_y^2+p_z^2}\right),$$

\noindent i.e. $m\rightarrow 0$.  This is analogous to `p-scheme' (lepton collider) jet clustering~\cite{peskin}, except with the mass constraint imposed as a post-prossessing step instead of a pre-processing step (the `E-scheme' is still used for the clustering).  The variable was first studied in hadron colliders more recently in the context of analyzing the jet energy scale uncertainty from the $W\rightarrow qq'$ boson mass~\cite{tancredi}.   The spatial mass has the key property that it is insensitive to the scale and resolution of the input object masses.  In the case that the two input objects are jets, this also means that $m_s$ is insensitive to the uncertainties associated with the calibrations of the input jet masses.  It is quite easy to generalize $m_s$ to the case of three or more input objects - simply apply $g$ to every input four-vector before computing the square.  The key consideration concerning $m_s$ is if the loss of information with respect to the full invariant mass is compensated by the reduction in uncertainty from the mass calibration.

\section{Top Mass Measurement with Small Radius Jets}

\begin{figure}[h!]
\begin{center}
\includegraphics[width=0.5\textwidth]{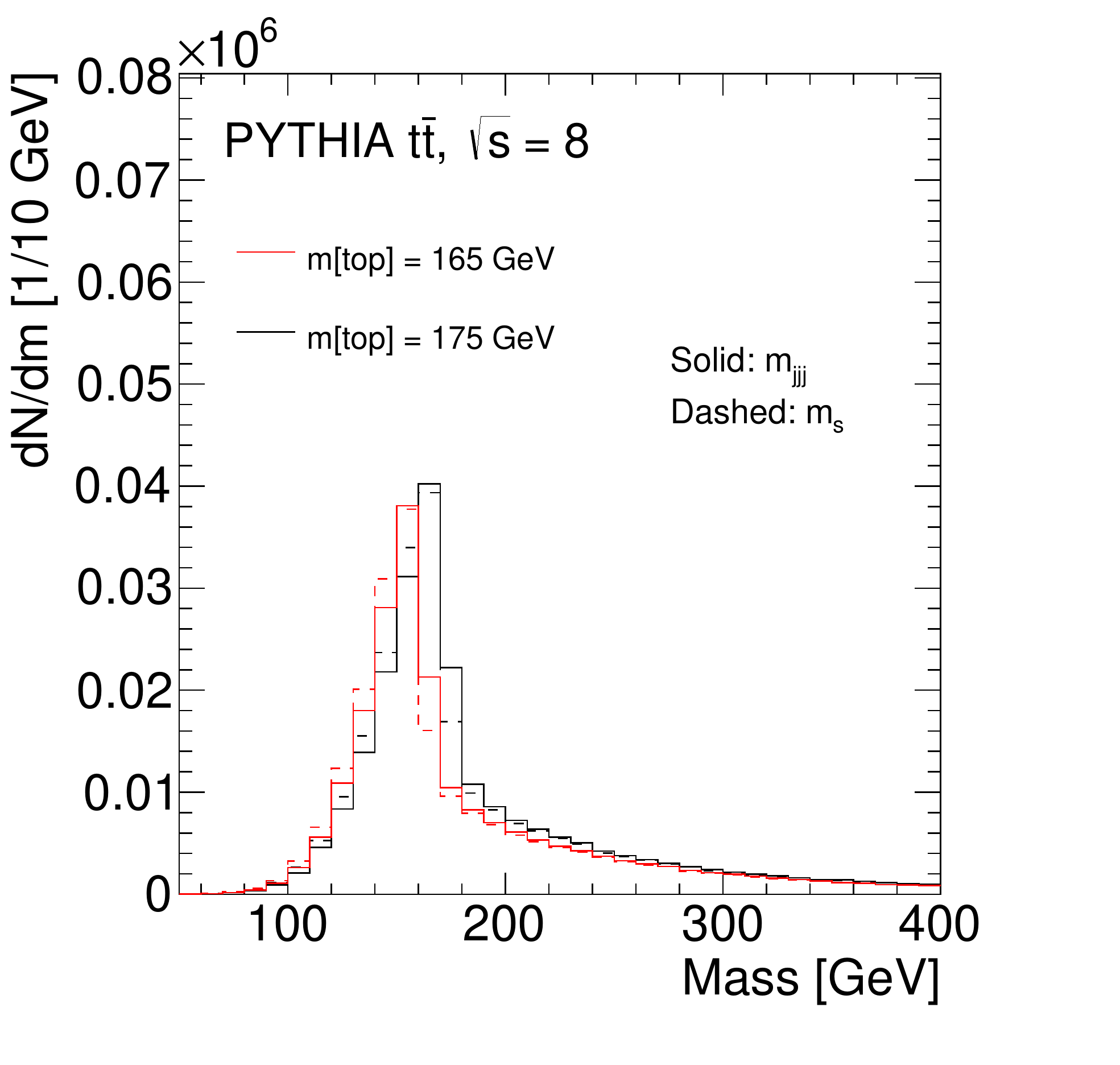}
\end{center}
\caption{The distribution of $m_{jjj}$ for various generated top quark masses.}
\label{fig:algo}
\end{figure}

One measurement that may be particularly sensitive to small radius jet masses is the determination of the top quark mass.  Most methods for measuring the top quark mass use a kinematic reconstruction of the decay products.  We will consider here a simplified top mass measurement in order to illustrate the effect of the small radius jet mass.  The measurement uses templates of a variable, $m_{jjj}$ which is sensitive to the top quark mass.  Top quark pair production ($t\bar{t}$) events are simulated using \textsc{Pythia} 6.425~\cite{Pythia} at $\sqrt{s}=8$ TeV with one interaction per $pp$ collision.  Rivet 1.8.2~\cite{rivet} is used as a framework to reconstruct events and select single lepton final states in order to identify $t\bar{t}$ decays where one of the $W$ bosons from the $t\rightarrow bW$ decays into leptons and the other decays hadronically.  Stable particles (excluding electrons and muons) with $p_T>500$ MeV and $|\eta|<5$ are clustered into jets using the \textsc{FastJet}~\cite{fastjet} 3.0.3 implementation of the anti-$k_t$~\cite{antikt} algorithm with $R=0.4$.  Jets are $b$-tagged by identifying $b$-hadrons from the Monte Carlo (MC) truth record within a $\Delta R=\sqrt{\Delta\phi^2+\Delta\eta^2}$ cone of $0.4$ of the jet axis.  We require at least four jets with $p_T>25$ GeV and at least two must be $b$-tagged.  Leptons are required to have $p_T>25$ GeV and be at least $\Delta R>0.4$ from any jet.  The missing transverse momentum is the negative of the vector sum of all stable particles within $|\eta|<5$.  Three jets $j_1,j_2,b_1$, exactly one with a $b$-tag ($b_1$), are associated with the hadronically decaying top quark by minimizing the following $\chi^2$:  

	$$\chi^2 = \frac{(m_{j_{1}j_{2}b_{1}}-m_{b_2l\nu})^2}{(20\text{ GeV})^2}+\frac{(m_{j_1 j_2}-m_W)^2}{(10\text{ GeV})^2},$$

\noindent where $j_i$ are from the set of all non $b$-tagged jets with $p_T>25$ GeV, $b_1$ and $b_2$ are the highest $p_T$ $b$-tagged jets (not necessarily in order), $m_W\sim 80$ GeV~\cite{pdg}, and the neutrino four-vector is determined from the missing transverse momentum in the $x$ and $y$ coordinates and by requiring $m_{l\nu}=m_W$ for the $z$ component\footnote{The solution to $m_{l\nu}=m_W$ is quadratic in the neutrino $p_z$ and the value corresponding to the smaller $\chi^2$ is used.  In some cases, there is no solution to the quadratic equation in which case the neutrino $p_z$ is set to zero.  The neutrino is assumed massless.}.  The template variable $m_{jjj}$ is simply $m_{j_1j_2b_1}$.  Figure~\ref{fig:algo} shows the distribution of $m_{jjj}$ for two generated top quark masses.  One way to measure the top quark mass is to perform a fit\footnote{Computed over a fixed range, which we take to be $100<m_{jjj}/\text{GeV}<200$.} to $\langle m_{jjj}\rangle$, for various templates like those shown in Fig.~\ref{fig:algo}.  

More sophisticated methods based on the full shapes of (multiple) distributions have been used to measure the top quark mass as well as simultaneously constrain (i.e. reduce) the jet energy scale uncertainty (see e.g.~\cite{bestCMS,atlastopmass}).  Such techniques, which take advantage of a $W$ boson mass constraint, can also contrain jet mass scale uncertainties, but not resolution uncertainties.  Such details are beyond the scope of this paper - we aim to illustrate the impact of small radius jet mass uncertainties in a simple case to demonstrate that there is a tradeoff between sensitivity and overall systematic uncertainty when comparing spatial mass with variables based on the full small radius jet four vectors.

\section{Impact of Small Radius Jet Uncertainties}

\begin{figure}[h!]
\begin{center}
\includegraphics[width=0.5\textwidth]{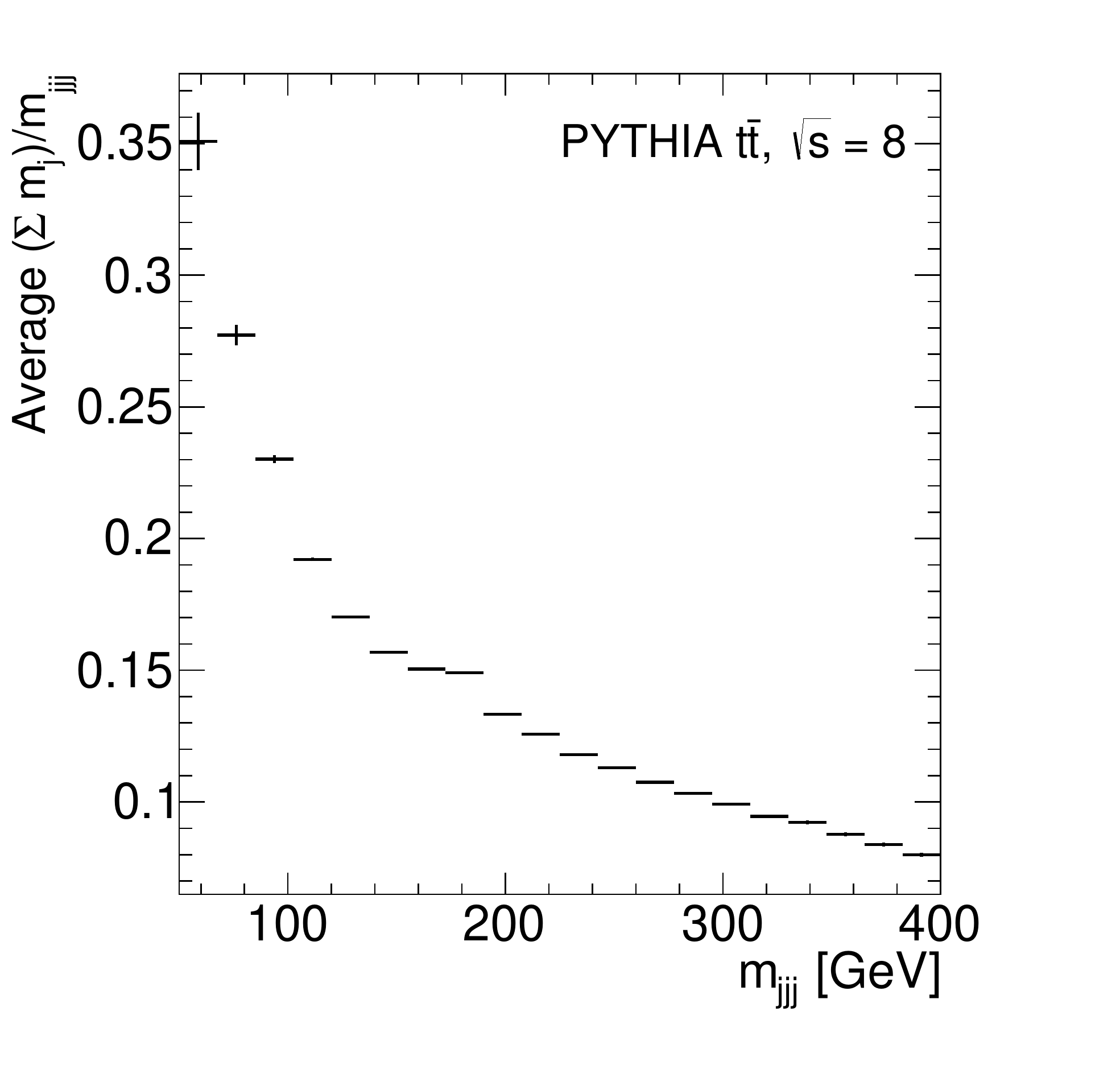}\includegraphics[width=0.5\textwidth]{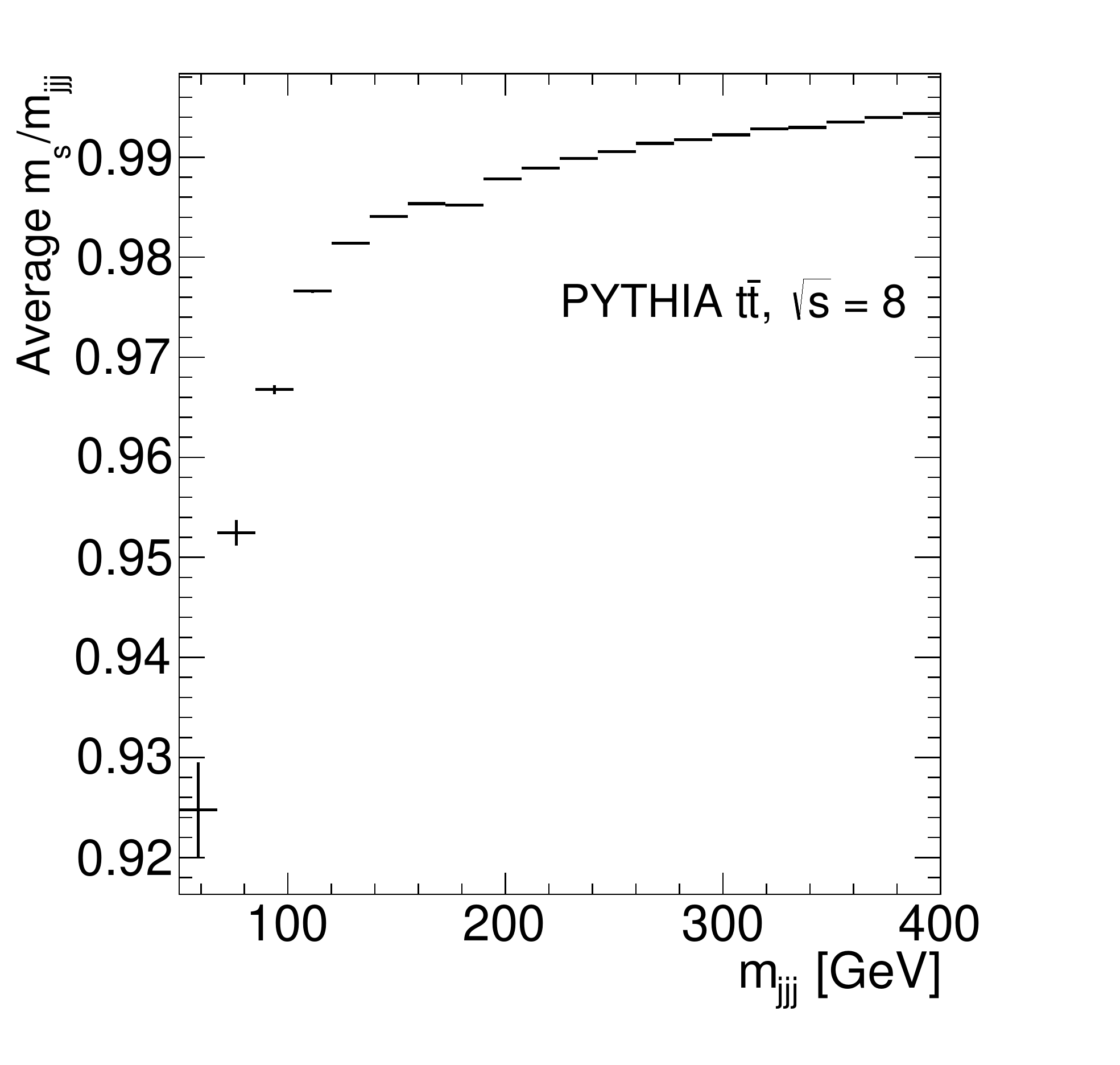}
\end{center}
\caption{Left: A measure of the contribution of the small radius jet mass to $m_{jjj}$ in the form of the ratio of the sum of the $W$ daughter jet masses and the matched $b$-jet mass with the trijet mass $m_{jjj}$ as a function of $m_{jjj}$.  Right: The ratio of $m_s$ to $m_{jjj}$ as a function of $m_{jjj}$. }
\label{fig:algo2}
\end{figure}

Let us now consider what impact the individual jet masses have on $m_{jjj}$ as described in the previous section.  The left plot of Fig.~\ref{fig:algo2} gives a measure of the contribution of the small radius jet masses to $m_{jjj}$ as a ratio of the scalar sum of the $W$ daughter jet masses and the mass of the associated $b$-jet to the total trijet invariant mass.  For a wide range of invariant masses, the contribution from the sum of the small radius jet masses is between 10-20\%.  The rest of the mass comes from the small radius jet $p_T$ and opening angles, which is captured by $m_s$ as shown in the right plot of Fig.~\ref{fig:algo2}.  The value of $m_s$ is well over 90\% of $m_{jjj}$ on average.

\begin{figure}[h!]
\begin{center}
\includegraphics[width=0.5\textwidth]{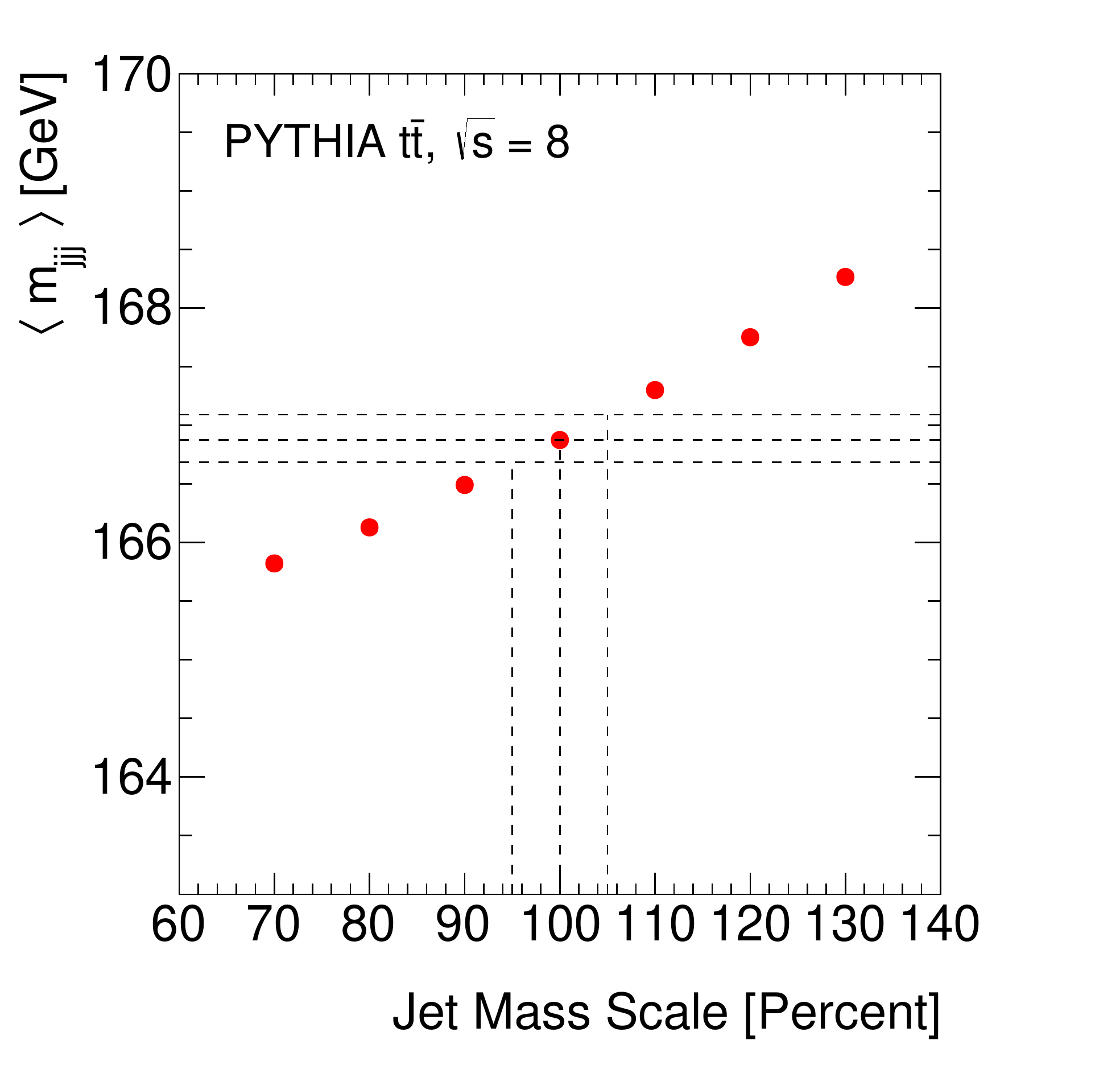}
\end{center}
\caption{The dependence of $\langle m_{jjj}\rangle $ on the small radius jet mass scale. 100\% corresponds to the nominal energy scale with a 30\% uncertainty corresponding to shifts between 70\% and 130\%.  The dashed lines mark the 5\% uncertainty level which is $\sim 200$ MeV on $\langle m_{jjj}\rangle$.}
\label{fig:algo3}
\end{figure}

Let us assume that the small radius jet masses can be calibrated to have zero bias, i.e. reconstructed jet masses are on average equal to the corresponding particle jet masses in simulations of the LHC detectors.  Then an important question to ask is, given that the small radius jet masses can contribute 10-20\% of the total invariant mass, how does an uncertainty on the small radius jet calibration translate to an uncertainty on $m_{jjj}$?  For instance, a systematic error in the mass scale calibration would correspond to a coherent shift of the three small radius jet masses.  Answering this question will give us an indication of how much we can gain in uncertainty reduction on the top mass by using $m_s$ over the trijet invariant mass. In principle, there are uncertainties on every moment of the small radius jet mass distribution.  For this illustration, we only consider the uncertainty on the first moment: the jet mass scale uncertainty.  Figure~\ref{fig:algo3} shows how $m_{jjj}$ depends on a scale shift of the small radius jet masses.  The point 100\% corresponds to the nominal $\langle m_{jjj}\rangle $ at $m_{top}=175$ GeV.  Large radius jets have been shown to have mass scale uncertainties around 5\%~\cite{atlaslargeR}; at the same level of uncertainty for small radius jets (compare the shift between 95\% and 105\% in Fig.~\ref{fig:algo3}), the change in $\langle m_{jjj}\rangle$ is about 200 MeV which corresponds to a $\Delta m_\text{top}\sim$ 400 MeV according to the calibration curve in Fig.~\ref{fig:algo4} and is comparable to the dominant systematic uncertainties in the single most precise top quark mass measurement recently announced by CMS~\cite{bestCMS}.  However, if the uncertainty on the small radius jet mass is 1\% instead of 5\%, then the propagated uncertainty of $\lesssim 80$ MeV becomes subdominant. This illustrates the importance of testing the tradeoff between the reduced uncertainty of $m_s$ and the loss of information with respect to the full invariant mass.  The next section quantifies the `information' contained in the invariant mass.

\section{Spatial Mass versus Invariant Mass}

\begin{figure}[h!]
\begin{center}
\includegraphics[width=0.5\textwidth]{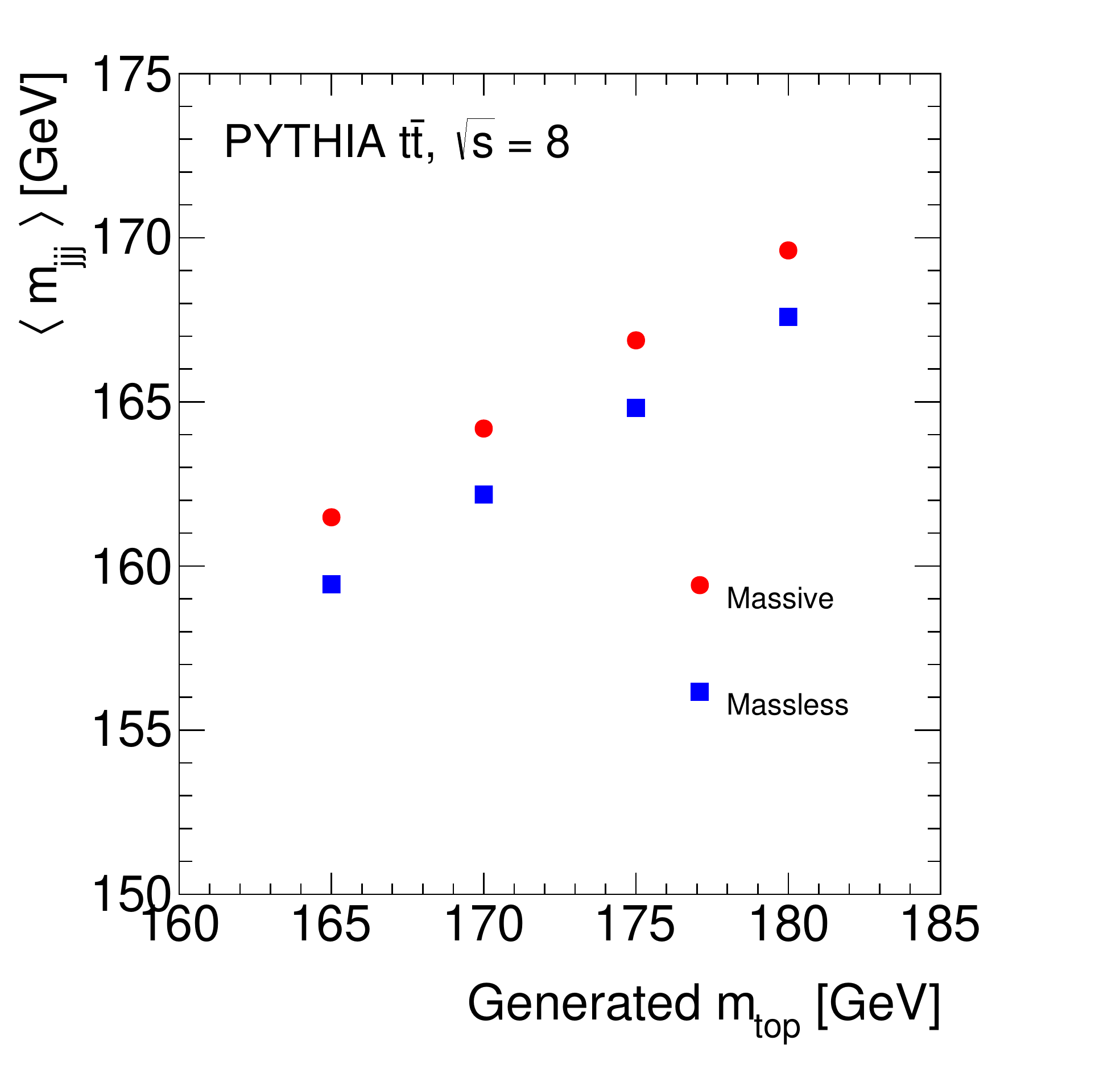}
\end{center}
\caption{The dependance of $\langle m_{jjj}\rangle$ and $\langle m_s \rangle$ on $m_{top}$.  Using these curves, we can translate between uncertainties on $\langle m_{jjj}\rangle$ or $\langle m_s\rangle$ and $m_\text{top}$ e.g. 200 MeV for a $\sim 5\%$ uncertainty on $\langle m_{jjj}\rangle$ translates to a $\sim 400$ MeV uncertainty in $m_\text{top}$.}
\label{fig:algo4}
\end{figure}

There are several possible methods for mitigating the impact of the small radius jet mass uncertainties.  Since the term in the uncertainty of the invariant mass scales with $m_j/m_{jjj}$, one method is to require that $m_j\ll m_{jjj}$ for all jets.  This has been employed in a recent mass measurement by CMS using kinematic endpoints~\cite{cms}.  However, this approach reduces the signal acceptance.  The spatial mass, on the other hand, is insensitive to the jet mass uncertainties without effecting the selection efficiency.   Since the small radius jet masses do carry a non-trivial fraction of $m_{jjj}$ as was observed in Fig.~\ref{fig:algo2}, the scale of the $m_{jjj}$ distribution shifts when using $m_s$ instead of $m_{jjj}$. However, the relevant quantity is not the scale of the invariant mass (which can be calibrated using Monte Carlo simulation), but how the invariant mass changes with $m_{top}$.  The `mass determination power' in a template method based on a variable $x$ is determined by $\beta_x\equiv \partial x/\partial m_{top}$: the more the variable changes with $m_{top}$, the more sensitive a likelihood-based method can accurately determine the true value of the top quark mass.  Figure~\ref{fig:algo4} shows how $\langle m_{jjj}\rangle$ and $\langle m_s\rangle$ scale with the generated $m_{top}$.  It is clear that there is a scale shift between $\langle m_{jjj}\rangle(m_{top})$ and $\langle m_s\rangle(m_{top})$.  However, the slopes are nearly identical.  Quantitatively, a linear fit yields a slope of $\beta_{m_{jjj}}=0.542\pm 0.006$ for the massive small radius jets and $\beta_{m_s}=0.541\pm 0.006$ for spatial mass - consistent within statistical uncertainty.   The current systematic uncertainty on the top quark mass are on the scale of  $\Delta_\text{top}=$ 1-2 GeV.   Using $\beta_{m_{jjj}}$, we can translate this into an uncertainty on $\langle m_{jjj}\rangle$ and then translate back to $m_\text{top}$ using $\beta_{m_s}^{-1}$ to see how the 1-2 GeV changes with a different calibration:

\begin{align*}
\Delta_\text{top}^{m_s}\sim \beta_{m_s}^{-1} \Delta \langle m_s\rangle \sim \beta_{m_s}^{-1} \Delta\langle m_{jjj}\rangle \sim \beta_{m_s}^{-1}\beta_{m_{jjj}} \Delta_\text{top}^{m_{jjj}}\sim (\beta_{m_s}^{-1}\beta_{m_{jjj}})\times \text{1-2 GeV.}
\end{align*}

\vspace{2mm}

\noindent Two calibration curves with identical slope will give the same translation between the template and the top quark mass.  A calibration curve with a smaller slope will increase the uncertainty on $m_\text{top}$ (as $\beta_{m_s}\rightarrow 0, \Delta_\text{top}\rightarrow\infty$).   In this particular case, the use of $m_{s}$ instead of $m_{jjj}$ magnifies the 1-2 GeV by 2-4 MeV when using the nominal values (0.542 and 0.541) for the slopes, though the slopes are consistent with zero within uncertainties.  This is negligible compared to the potential several hundred MeV reduction from the lack of small radius jet mass uncertainty in $m_s$ over $m_{jjj}$. 

In addition to the experimental uncertainties associated with jet mass calibration, it is important to check that the spatial mass method has a smaller or comparable theoretical modeling uncertainties.  For this simplified mass measurement, the difference between e.g. Pythia and Herwig (version 6.510~\cite{herwig}) is large, but the same difference is observed for $m_{jjj}$ and $m_s$ and thus the two methods approximate the same uncertainty.


\section{Conclusion}

Many searches and measurements at the LHC use the full four vector of small radius jets.  We have shown that replacing the invariant mass with the spatial mass, a variable constructed from only the spatial components of the input jet four vectors, provides a method for eliminating the impact of the small radius jet mass uncertainty.  Furthermore, in the example shown here, we find that the performance of $m_s$ is comparable to $m_{jjj}$ which means that the overall precision has improved.  This may not be true for all cases, but the illustrative top mass example suggests that analyses at the LHC should assess the impact of small radius jet masses on their results and consider the use of $m_s$ as a means for removing the dependance on the corresponding uncertainties.

\section{Acknowledgments}

We would like to specially thank Michael Peskin, Steve Ellis, and Aurelio Juste for very useful discussions and detailed comments and suggestions on the text. We are also grateful to Maximilian Swiatlowski and Tancredi Carli for many interesting conversations.  This work is supported by the US Department of Energy (DOE) Early Career Research Program and grant DE-AC02-76SF00515.  BN is supported by NSF Graduate Research Fellowship under Grant No. DGE-4747 and by the Stanford Graduate Fellowship.

\end{document}